\documentclass[twocolumn,aps,prl,showpacs,superscriptaddress]{revtex4}
\usepackage{color}
\usepackage{mathtools}
\usepackage{graphicx}
\usepackage{SIunits}

\begin{document}

\title{{Andreev Reflection Spectroscopy of Topological Superconductor Candidate Nb$_x$Bi$_2$Se$_3$}}

\author{C.~Kurter}
\affiliation{Department of Physics and Materials Research Center, Missouri University of Science and Technology, Rolla, MO 65409}
\affiliation{Department of Physics and Materials Research Laboratory, University of Illinois at Urbana-Champaign,
Urbana, IL 61801}

\author{A.~D.~K.~Finck}
\affiliation{Department of Physics and Materials Research Laboratory, University of Illinois at Urbana-Champaign,
Urbana, IL 61801}

\author{E.~D.~Huemiller}
\affiliation{Department of Physics and Materials Research Laboratory, University of Illinois at Urbana-Champaign,
Urbana, IL 61801}

\author{J.~Medvedeva}
\affiliation{Department of Physics and Materials Research Center, Missouri University of Science and Technology, Rolla, MO 65409}

\author{A.~Weis}
\affiliation{Department of Physics and Materials Research Laboratory, University of Illinois at Urbana-Champaign,
Urbana, IL 61801}

\author{J.~M.~Atkinson}
\affiliation{Department of Physics and Materials Research Laboratory, University of Illinois at Urbana-Champaign,
Urbana, IL 61801}

\author{Y.~Qiu}
\affiliation{Department of Physics and Materials Research Center, Missouri University of Science and Technology, Rolla, MO 65409}
\author{L.~Shen}
\affiliation{Department of Physics and Materials Research Center, Missouri University of Science and Technology, Rolla, MO 65409}
\author{S.~H.~Lee}
\affiliation{Department of Physics and Materials Research Center, Missouri University of Science and Technology, Rolla, MO 65409}
\author{T.~Vojta}
\affiliation{Department of Physics and Materials Research Center, Missouri University of Science and Technology, Rolla, MO 65409}

\author{P. Ghaemi}
\affiliation{Department of Physics, City College of New of CUNY, New York, NY 10031}
\affiliation{Department of Physics, Graduate Center of CUNY, New York, NY 10016}

\author{Y.~S.~Hor}
\affiliation{Department of Physics and Materials Research Center, Missouri University of Science and Technology, Rolla, MO 65409}

\author{D.~J.~Van Harlingen}
\affiliation{Department of Physics and Materials Research Laboratory, University of Illinois at Urbana-Champaign,
Urbana, IL 61801}

\date{\today}

\begin{abstract}

We study unconventional superconductivity in thin exfoliated single crystals of a promising 3D topological superconductor candidate, Nb-doped Bi$_2$Se$_3$ through Andreev reflection spectroscopy and magneto-transport.  Measurements of Andreev reflection in low and high resistance samples both show enhanced conductance around zero bias and conductance dips at the superconducting energy gap. Such behavior is inconsistent with conventional Blonder-Tinkham-Klapwijk theory of Andreev reflection. We discuss how our results are consistent with $p$-wave pairing symmetry, supporting the possibility of topological superconductivity in Nb-doped Bi$_2$Se$_3$.
\end{abstract}


\maketitle


There has been tremendous effort to realize topological superconductivity in three dimensional topological insulators either through proximity coupling to superconductors~\cite{NatCommun.2.575, PhysRevLett.109.056803, NatMat.11.417, NatCommun.4.1689, PhysRevB.90.014501, PhysRevB.93.035307, KurterNature} or through chemical dopants and intercalations~\cite{PhysRevLett.104.057001,PhysRevLett.107.217001,YewSanArXIV, doi:10.1021/jacs.5b06815}. These superconductors are speculated to possess an unconventional superconducting gap with $p$-wave pairing symmetry~\cite{PhysRevB.88.024515,PhysRevLett.108.107005} that generally results in the formation of subgap surface Andreev bound states. Such exotic superconductors host zero energy Bogoliubov quasiparticles that mimic Majorana fermions~\cite{Majorana,NatPhys.5.614}. Majorana modes have a great potential to implement topologically protected quantum computation~\cite{RevModPhys.80.1083}.


One common method to investigate an exotic superconductor is to perform Andreev reflection spectroscopy on a localized normal metal-superconductor (NS) junction~\cite{Tinkham}. This method has been employed to study conductance spectra obtained with a gold contact on superconducting doped topological insulators such as Cu$_x$Bi$_2$Se$_3$. The robust zero bias peaks at low temperatures are attributed to zero energy surface bound states and $p$-wave pairing symmetry~\cite{PhysRevLett.107.217001}. These intriguing results stimulated further spectroscopic experiments. However, low temperature scanning tunneling microscopy measurements exhibited contradictory results~\cite{PhysRevLett.110.117001} that were consistent with $s$-wave pairing. Recent nuclear magnetic resonance experiments presented more conclusive findings by showing spontaneously broken spin-rotation symmetry in the hexagonal plane of the electron-doped topological insulator Cu$_{0.3}$Bi$_2$Se$_3$ in the superconducting state suggesting evidence for spin-triplet pairing and nematicity~\cite{NatPhys12_852}.


The devices we study in this paper employ localized normal metal leads fabricated on top of thin flakes of Nb-doped topological insulator Bi$_2$Se$_3$. The material is claimed to have coexistence of magnetic ordering and unusual superconductivity~\cite{YewSanArXIV} due to spontaneous time reversal symmetry breaking~\cite{ChirolliArXIV, LawArXIV}. Moreover, torque magnetometry measurements reported rotational symmetry breaking suggesting a nematic order in the superconducting ground state of Nb-doped Bi$_2$Se$_3$, consistent with an $E_u$ pairing symmetry~\cite{PhysRevX.7.011009}. Complementary directional penetration depth measurements supported the presence of rotational symmetry breaking and a nodal superconducting energy gap~\cite{PhysRevB.94.180510}.  Here, we report anomalous Andreev reflection spectra that are inconsistent with conventional Blonder-Tinkham-Klapwijk theory for $s$-wave superconductivity \cite{PhysRevB.25.4515}.  These features, including zero bias conductance peaks and conductance dips at the superconducting pairing energy, can be described by certain forms of $p$-wave superconductivity.

\begin{figure}
\centering
\includegraphics[bb=1 1 719 323,width= 3.5 in]{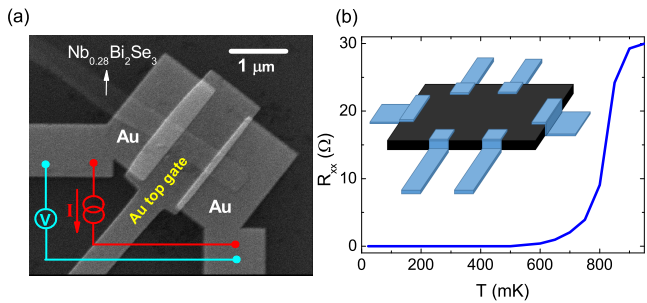}
\caption{(Color online) (a)~Scanning electron microscopy image of a representative device made with two gold leads on an exfoliated piece of Nb$_{0.28}$Bi$_2$Se$_3$. The configuration of current and voltage probes for 2-terminal measurements are schematically shown. (b) Longitudinal resistance as a function of temperature from a Hall bar device fabricated on a similar flake showing the transition temperature. Inset shows the schematics of the hall bar device.} \label{fig:SEM_v2}
\end{figure}

Fig.~\ref{fig:SEM_v2}(a) shows a scanning electron microscopy image of a representative device with two Au localized contacts on top of an exfoliated thin flake of Nb$_{0.28}$Bi$_2$Se$_3$. The bulk samples show robust superconductivity with a transition temperature of 3.5 K~\cite{YewSanArXIV}. The method by which high quality samples were grown is discussed in Ref~\cite{YewSanArXIV}. Starting with single crystals of Nb$_{0.28}$Bi$_2$Se$_3$, we transfer thin flakes onto doped Si substrates with a 300 nm thick oxide layer through the traditional scotch tape method. After thin, flat, and large area pieces are identified with atomic force microscopy, we define the junction leads by e-beam lithography and lift-off of an e-beam evaporated film of 50 nm Au.  Prior to metallization, we employ \emph{in situ} Ar ion-milling of the sample surface to remove any native oxides or other contaminants. The light etching (about a few seconds with a beam current of 10 mA and beam voltage of 400 V) of the crystal surface cleans the NS interface and improves contact resistance. Etching will also produce a rougher surface, with multiple crystallographic planes contributing to the NS interface. Incomplete etching will generate NS junctions with thin tunnel barriers with reduced transparency and higher resistances.  Although not used in the measurements presented here, a top gate was created by first depositing a thin layer of Al$_2$O$_3$ on the devices via atomic layer deposition. The top gate is defined with a second lithographic process (see Fig.~\ref{fig:SEM_v2}a), followed by e-beam evaporation of Au.  The devices were thermally anchored to the mixing chamber of a cryogen-free dilution refrigerator equipped with a vector magnet and filtered wiring. We perform low frequency transport measurements with standard lock-in techniques, typically with a 10 nA AC excitation at f = 73 Hz.

Devices with six contacts in a Hall bar geometry as shown in the inset of Fig.~\ref{fig:SEM_v2}(b) allow measurements of intrinsic dissipation $R_{xx}$ while avoiding the influence of contact resistance. The longitudinal resistance vs temperature obtained from a Hall bar device fabricated on a 17 nm thick flake demonstrates a transition temperature of 800 mK as shown in the main panel of Fig.~\ref{fig:SEM_v2}(b). This is significantly reduced from the bulk value due to either finite size effects or disorder. 

\begin{figure}
\centering
\includegraphics[bb=2 2 647 576,width= 3.5 in]{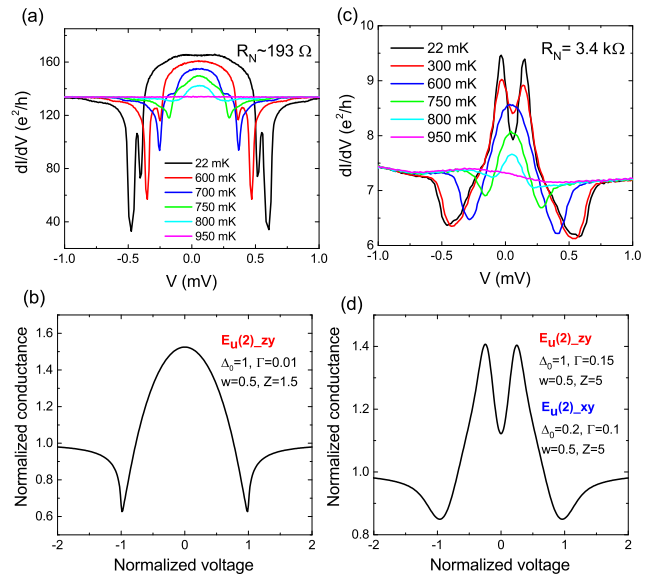}
\caption{(Color online) Measured Andreev reflection spectra at various temperatures for (a) a low resistance device with R$_N$= 193 ohms and (c) a high resistance device with R$_N$= 3.4 kohms.  Below, simulated spectra for (b) the low resistance device and (d) the high resistance device.  In (b) and (d), voltage is normalized by the superconducting energy gap and conductance is normalized by the normal state conductance, R$_N$.} \label{fig:Andreev}
\end{figure}

Devices with only two or three normal metal leads are employed for Andreev reflection spectroscopy of the NS interface.  Differential conductance vs.~voltage, dI/dV vs.~V, characteristics of a representative sample with relatively transparent Au/Nb$_{0.28}$Bi$_2$Se$_3$ interface are shown in Fig.~\ref{fig:Andreev}(a) for various temperatures. Below the critical temperature, the device shows significant enhancement in conductance around zero bias due to Andreev reflection at the interface of Nb$_{0.28}$Bi$_2$Se$_3$ and Au followed by additional high bias features. These broad zero bias conductance (ZBC) peaks are accompanied by sharp dips at the superconducting gap, which are not observed in superconductors with conventional $s$-wave pairing symmetry. In fact, this abrupt suppression of conductance at superconducting gap is a signature of shifting spectral weight away from high energy states to form low energy surface bound states and is often associated with $p$-wave pairing symmetry~\cite{PhysRevB.85.180509}.  We typically observe energy gaps in the range of $100-500$ $\mu$V. 

Although a finite critical current might also generate spurious ZBC peaks \cite{PhysRevB.69.134507}, we observe that above 300 mK the zero bias conductance continually drops as temperature increases (See the supplemental material, Fig.~S1a). Similarly, conductance at zero bias also decreases in the presence of a small magnetic field (Fig.~S1b).  As argued by Ref.~\cite{PhysRevLett.107.217001}, this behavior is incompatible with spurious zero bias conductance peaks generated by finite critical current, which would leave zero bias conductance unchanged after small rises in temperature or magnetic field.

Although there has been numerous calculations for the Andreev conductance spectra for the various categories of odd-parity and topological superconductivity (e.g. Refs.~\cite{PhysRevB.56.7847, KashiwayaJPSJ, PhysRevB.85.180509, PhysRevB.92.205424}), determining which pairing symmetry corresponds to an observed conductance spectra is problematic because the presence of multiple fitting parameters permit agreement with more than one model.  For example, in Figs.~\ref{fig:Andreev}(b) and 2(d), we show calculated conductance spectrum of a $p$-wave symmetry superconductor junction by using the $E_u(2)$ model shown in Ref~\cite{KashiwayaJPSJ} for moderate and high barrier strength, Z respectively. For such a symmetry, the matrix form of the pair potential, $\Delta$ can be written in terms of polar and azimuthal angles $\theta$ and $\phi$ where $\Delta_{\uparrow \uparrow }$($\theta$,$\phi$)= $\Delta_0 e^{i \phi} \sin{\theta}$ and $\Delta_{\uparrow \downarrow }$($\theta$,$\phi$)= $\Delta_{\downarrow \uparrow }$($\theta$,$\phi$)= $\Delta_{\downarrow \downarrow }$($\theta$,$\phi$)= 0. In that model, tunneling in the $z-y$ plane gives a prominent ZBC peak, as do several other odd-parity models in other directions. Because of the surface roughness induced by ion milling prior to deposition of the gold leads, it is likely that our devices include contributions from tunneling into both the $z-y$ and the $x-y$ planes of the Nb-doped Bi$_2$Se$_3$ crystal.  We do not presume $E_u(2)$ is the actual pairing symmetry of Nb$_{0.28}$Bi$_2$Se$_3$, but we select it for our modeling because other experimental studies are also consistent with this pairing symmetry \cite{PhysRevX.7.011009, PhysRevB.94.180510} and to show its agreement with our results.  While material-specific conditions such as interorbital interactions might alter the exact structure of the pairing interaction \cite{PhysRevB.95.201109, PhysRevB.95.201110}, it is probable that our simple formalism for the pairing should be sufficient to qualitatively match our measurements.  We choose a moderate barrier strength of $Z = 1.5$ for Fig.~\ref{fig:Andreev}(b) to represent the partial transparency of the NS interface and match the shape of the observed spectrum in Fig.~\ref{fig:Andreev}(a).

Having discussed the devices with relatively more transparent NS contacts, we now turn to conductance spectra of comparatively higher contact resistance devices. Such devices typically demonstrate a split peak at low temperatures instead of a single broad ZBC peak due to increasing barrier strength at the NS interface~\cite{PhysRevB.25.4515}. Fig.~\ref{fig:Andreev}(c) shows temperature dependence of differential conductance characteristics from a device with the normal state resistance R$_N$ of 3.4 k$\ohm$. The well-defined split peak feature observed at low temperatures gets suppressed with temperature and eventually evolves into a single peak around zero bias beyond 600 mK.  Beyond 950 mK, superconductivity is completely lost.

Furthermore, one can see the coherent conductance dips just beyond the superconducting gaps as observed in devices with more transparent contacts.  The appearance of such conductance dips in a high resistance sample helps to rule out finite critical currents as the source of the anomalous suppression in conductance at the superconducting energy gap.

While the conductance dips at higher energy is consistent with tunneling into the $z-y$ plane of a $p$-wave superconductor with $E_u(2)$ pairing symmetry, the split peak is expected for tunneling into the $x-y$ plane.  Because of surface roughness from ion milling and the fact that the gold leads are in contact with both the top and side surfaces of the Nb-doped Bi$_2$Se$_3$ crystal, it is not unreasonable to expect that Andreev spectroscopy would include contributions from tunneling into both the $x-y$ and $z-y$ planes.  For NS interfaces parallel to the $x-y$ plane, the $E_u(2)$ pairing symmetry \cite{PhysRevB.56.7847, KashiwayaJPSJ} yields Andreev spectra with conductance peaks near the superconducting energy gap, as we observe.  We plot a simulated spectrum in Fig.~\ref{fig:Andreev}(d) using the $E_u(2)$ model with a barrier strength of $Z=5$.   To capture the split peak feature of Fig.~\ref{fig:Andreev}(b), we include contributions from tunneling into both the $z-y$ plane (with normalized pairing energy $\Delta_0 = 1$) and the $x-y$ plane (with a smaller pairing energy $\Delta_0 = 0.2$).

We also note that the split peak shape in Fig.~\ref{fig:Andreev}(c) can result from surface Andreev bound states \cite{PhysRevLett.107.217001, PhysRevLett.108.107005}.  Such subgap states form on the surface of topological superconductors and lead to a characteristic split peak feature in spectroscopic measurements due to van Hove singularities in their density of states \cite{PhysRevLett.108.107005}.  A caveat is that Refs.~\cite{PhysRevLett.107.217001} and \cite{PhysRevLett.108.107005} considered the case of time-reversal symmetric topological superconductors, although it is possible that similar features are observable in topological superconductors with broken time-reversal symmetry.

\begin{figure}
\centering
\includegraphics[bb=2 2 866 434,width= 3.5 in]{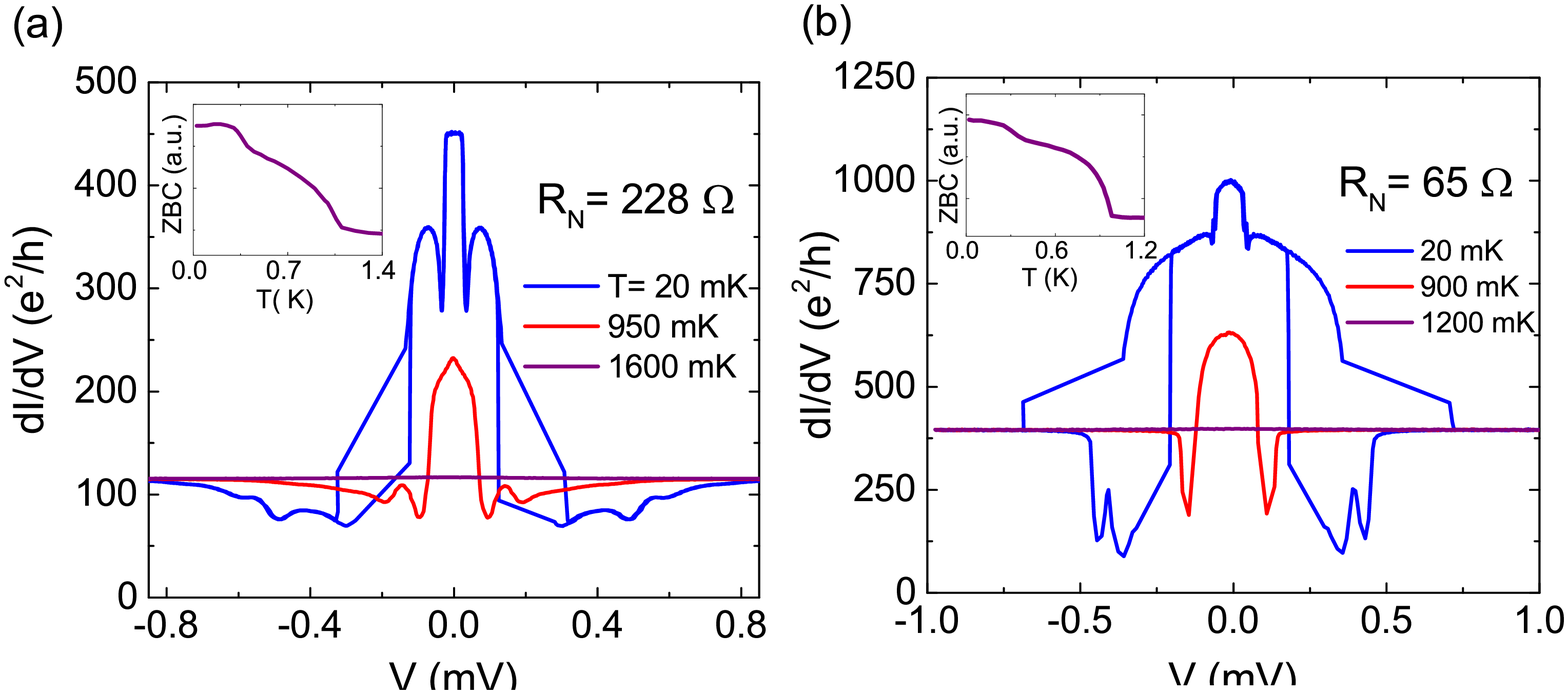}
\caption{(Color online) (a) and (b) Andreev spectra from two different devices that each contain two apparently different energy gaps.  Hysteretic current-voltage behavior leads to discontinuities in the measured spectra.} \label{fig:twogaps}
\end{figure}

In some of our samples, additional conductance features appear at the lowest temperatures.  In Fig.~\ref{fig:twogaps}(a) and (b), the 20 mK traces exhibit an additional ZBC peak that is superimposed on the broader main ZBC peak.  Like the main broad Andreev reflection feature, this additional conductance peak is accompanied by strong dips in conductance.  At the low temperatures, the observed I-V characteristics are highly hysteretic, possibly due to a circuit instability caused by a large negative differential conductance~\cite{PhysRevB.88.165308} or by heating; in any event the coherence dips associated with the main Andreev reflection features are partially obscured at 20 mK traces. As the temperature is increased but still below T$_c$, these dips are clearly observed in spectra (see the conductance spectra for each sample in red). The insets show the temperature dependence of the zero bias conductance vs temperature, in which two distinct transitions are apparent. The additional Andreev reflection features disappear roughly around 300 mK and suggest the presence of an extra smaller gap closing at lower temperatures. Hints of two gaps are also visible in the two sets of conductance dips in Fig.~\ref{fig:Andreev}(a), possibly corresponding to two gaps of similar but distinct magnitudes.  The nature of the multiple gaps is an open question and could originate from a non-trivial electronic band structure of Nb-doped Bi$_2$Se$_3$ as observed in other exotic materials such as Sr$_2$RuO$_4$~\cite{PhysRevLett.107.077003}. Our calculated band structures of Nb intercalated Bi$_2$Se$_3$ slab (See the supplemental material, Fig.~S2 and Fig.~S3) show that Nb states hybridize with host Bi$_2$Se$_3$ states near the Fermi level resulting in a spin-dependent multiple Fermi surfaces~\cite{PhysRevB.94.041114}, which may support our experimental observations of Andreev conductance spectra with multiple gap features.

The additional Andreev reflection features at low temperatures and low energies in Fig.~\ref{fig:twogaps} could also result from spatial inhomogeneity of the superconducting gap in the NS devices.  For example, there might be regions of weakened superconductivity in the Nb$_{0.28}$Bi$_2$Se$_3$ crystals, such as in the regions directly in contact with the gold probe, which could exhibit a smaller superconducting gap from an inverse proximity effect \cite{PhysRevX.4.011033}.  On the other hand, the inverse proximity effect is expected to result in a smoothly varying gap in the superconductor near the NS interface.  It is not clear to us how the inverse proximity effect might create a single, distinctly smaller gap observed through spectroscopy.

\begin{figure}
\centering
\includegraphics[bb=12 14 644 574,width= 3.5 in]{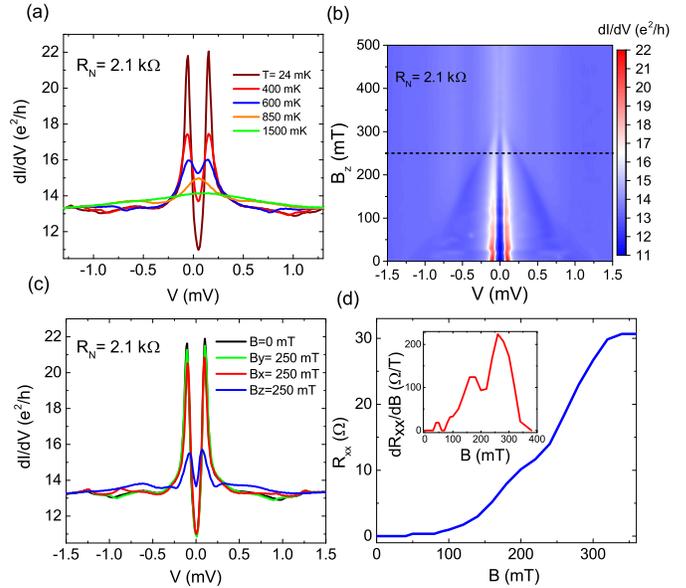}
\caption{(Color online) (a) Temperature dependence of differential conductance of one of the high resistance devices. (b) Color plot of out of magnetic field dependence of dI/dV for the same device. (c) Comparison of in-plane and out-of-plane magnetic field dependence of differential conductance, showing a strong directional anistropy.(d) Longitudinal resistance as a function of out-of-plane magnetic field obtained from a Hall bar device.} \label{fig:magnetic_field_v2}
\end{figure}

We finally discuss the magnetotransport properties of our devices. We first consider another high resistance Andreev reflection device with R$_N$= 2.1 k$\ohm$ whose temperature dependent conductance spectra at zero field is shown in Fig.~\ref{fig:magnetic_field_v2}(a).  Similar to the device shown in Fig.~\ref{fig:Andreev}(c), here we also observe a split peak centered on zero bias with depressed conductance at finite bias.  We plot the evolution of the Andreev spectrum of the device with out of plane magnetic field, $B_z$, in Fig.~\ref{fig:magnetic_field_v2}(b). Curiously, the location of the two peaks remain essentially independent of magnetic field.  Faint signatures of the two peaks remain even after the magnetic field has fully quenched other features of Andreev reflection.  We do not observe a splitting of the peaks with magnetic field, helping to rule out a Kondo effect \cite{nature.391.156} or other types magnetic anomalies as an origin for the zero bias conductance peaks.

The magnetotransport measurements of bulk crystals of the Nb-doped Bi$_2$Se$_3$ have been reported to show strong directional anisotropy \cite{PhysRevX.7.011009}. Thus, we explore the magnetic field direction anisotropy on conductance characteristics on thin flakes. Fig.~\ref{fig:magnetic_field_v2}(c) show differential conductance traces of the same high resistance sample at 250 mT as the magnetic field is applied in-plane (B$_x$ or B$_y$) and out-of-plane (B$_z$) orientations with respect to the substrate. The blue curve is the single conductance trace with perpendicular magnetic field marked with dashed lines in Fig.~\ref{fig:magnetic_field_v2}(b). The conductance spectra are highly affected by the out-of-plane magnetic field but remain almost unchanged with the in-plane magnetic field. All spectral features are suppressed beyond 300 mT with out-of-plane magnetic field as seen in Fig.~\ref{fig:magnetic_field_v2}(b). This suggests that there is a strong anisotropy of critical field between in-plane and out-of-plane directions.  Thus, our thin flakes are in a quasi two-dimensional limit, in which orbital effects are determined by mostly in-plane motion rather than out-of-plane motion. It is not yet clear whether this reduced dimensionality arises from the extreme thinness of our devices compared to the superconducting coherence length or reflects the dominance of two-dimensional surface states.

Fig.~\ref{fig:magnetic_field_v2}(d) shows longitudinal resistance as a function of out-of-plane magnetic field obtained from the same Hall bar device in Fig.~\ref{fig:SEM_v2}(b) that is constructed from a 17 nm thick flake. Prior to reaching the upper critical magnetic field, we observe finite resistance, suggesting the dissipative flow of magnetic vortices \cite{PhysRev.139.A1163}.  The resistance begins to rise most dramatically beyond 100 mT, which we tentatively identify as the lower critical field for this particular thin flake.  One can observe at least two distinct transitions with the magnetic field, which are absent in the temperature dependence of resistance shown in Fig.~\ref{fig:SEM_v2}(b). The derivation of R$_{xx}$ with respect to magnetic field in the insets of Fig.~\ref{fig:magnetic_field_v2}(d) gives three peaks corresponding to these transitions. These multiple critical fields suggest the formation of different vortex states \cite{Nature.390.259}.  Our results suggest the feasibility of introducing magnetic vortices in Nb-doped Bi$_2$Se$_3$, which could potentially harbor zero-energy Majorana bound states \cite{PhysRevLett.100.096407, PhysRevLett.107.097001}.

In summary, we explored Andreev reflection spectroscopy of exfoliated pieces of Nb-doped Bi$_2$Se$_3$. At low temperatures, we observed a broad enhancement in conductance at low energies, consistent with Andreev reflection.  However, the enhanced conductance persists even in high resistance devices and are accompanied by coherence dips at the superconducting gap that are not explained with conventional BTK theory and are associated with $p$-wave symmetry pairing.  Although these conductance spectra are consistent with $p$-wave superconductors, further directional tunneling and phase-sensitive Josephson interferometry measurements will help to confirm this. We also observed upper critical field anisotropy between in-plane and out-of-plane directions suggesting the flakes are in the two-dimensional limit. Finally, we observed multiple gap features --- tunneling spectroscopy measurements will be needed to determine if these arise from intrinsic two-gap superconductivity in the crystal or from spatially-separated regions with inhomogeneous superconducting properties.

$\it{Acknowledgements.}$ CK, ADKF, EDH, and DJVH acknowledge funding by Microsoft Station-Q. EDH and DJVH acknowledge funding from the National Science Foundation grant number DMR-1411067. TV acknowledges funding from the National Science Foundation grant number DMR-1506152. PG acknowledges support from the National Science Foundation CREST Center for Interface Design and Engineered Assembly of Low Dimensional systems (IDEALS) and the National Science Foundation grant number HRD-1547830.  YSH acknowledges funding from the National Science Foundation grant number DMR-1255607. For the device fabrication, we acknowledge use of the facilities of the Frederick Seitz Materials Research Laboratory at the University of Illinois at Urbana-Champaign, supported by the Department of Energy. We would like to thank Jessica Terbush for assistance of imaging the samples. We are thankful for helpful discussions with Fan Zhang.

\bibliography{NbxBi2Se3}

\begin{thebibliography}{5}
\expandafter\ifx\csname natexlab\endcsname\relax\def\natexlab#1{#1}\fi
\expandafter\ifx\csname bibnamefont\endcsname\relax
  \def\bibnamefont#1{#1}\fi
\expandafter\ifx\csname bibfnamefont\endcsname\relax
  \def\bibfnamefont#1{#1}\fi
\expandafter\ifx\csname citenamefont\endcsname\relax
  \def\citenamefont#1{#1}\fi
\expandafter\ifx\csname url\endcsname\relax
  \def\url#1{\texttt{#1}}\fi
\expandafter\ifx\csname urlprefix\endcsname\relax\def\urlprefix{URL }\fi
\providecommand{\bibinfo}[2]{#2}
\providecommand{\eprint}[2][]{\url{#2}}

\bibitem[{\citenamefont{Sasaki et~al.}(2011)\citenamefont{Sasaki, Kriener,
  Segawa, Yada, Tanaka, Sato, and Ando}}]{PhysRevLett.107.217001}
\bibinfo{author}{\bibfnamefont{S.}~\bibnamefont{Sasaki}},
  \bibinfo{author}{\bibfnamefont{M.}~\bibnamefont{Kriener}},
  \bibinfo{author}{\bibfnamefont{K.}~\bibnamefont{Segawa}},
  \bibinfo{author}{\bibfnamefont{K.}~\bibnamefont{Yada}},
  \bibinfo{author}{\bibfnamefont{Y.}~\bibnamefont{Tanaka}},
  \bibinfo{author}{\bibfnamefont{M.}~\bibnamefont{Sato}}, \bibnamefont{and}
  \bibinfo{author}{\bibfnamefont{Y.}~\bibnamefont{Ando}},
  \bibinfo{journal}{Phys. Rev. Lett.} \textbf{\bibinfo{volume}{107}},
  \bibinfo{pages}{217001} (\bibinfo{year}{2011}),
  \urlprefix\url{http://link.aps.org/doi/10.1103/PhysRevLett.107.217001}.

\bibitem[{\citenamefont{Perdew et~al.}(1996)\citenamefont{Perdew, Burke, and
  Ernzerhof}}]{PhysRevLett.77.3865}
\bibinfo{author}{\bibfnamefont{J.~P.} \bibnamefont{Perdew}},
  \bibinfo{author}{\bibfnamefont{K.}~\bibnamefont{Burke}}, \bibnamefont{and}
  \bibinfo{author}{\bibfnamefont{M.}~\bibnamefont{Ernzerhof}},
  \bibinfo{journal}{Phys. Rev. Lett.} \textbf{\bibinfo{volume}{77}},
  \bibinfo{pages}{3865} (\bibinfo{year}{1996}),
  \urlprefix\url{http://link.aps.org/doi/10.1103/PhysRevLett.77.3865}.

\bibitem[{\citenamefont{Perdew et~al.}(1997)\citenamefont{Perdew, Burke, and
  Ernzerhof}}]{PhysRevLett.78.1396}
\bibinfo{author}{\bibfnamefont{J.~P.} \bibnamefont{Perdew}},
  \bibinfo{author}{\bibfnamefont{K.}~\bibnamefont{Burke}}, \bibnamefont{and}
  \bibinfo{author}{\bibfnamefont{M.}~\bibnamefont{Ernzerhof}},
  \bibinfo{journal}{Phys. Rev. Lett.} \textbf{\bibinfo{volume}{78}},
  \bibinfo{pages}{1396} (\bibinfo{year}{1997}),
  \urlprefix\url{http://link.aps.org/doi/10.1103/PhysRevLett.78.1396}.

\bibitem[{\citenamefont{Qiu et~al.}(2016)\citenamefont{Qiu, Sanders, Dai,
  Medvedeva, Wu, Ghaemi, Vojta, and Hor}}]{YewSanArXIV}
\bibinfo{author}{\bibfnamefont{Y.}~\bibnamefont{Qiu}},
  \bibinfo{author}{\bibfnamefont{K.~N.} \bibnamefont{Sanders}},
  \bibinfo{author}{\bibfnamefont{J.}~\bibnamefont{Dai}},
  \bibinfo{author}{\bibfnamefont{J.~E.} \bibnamefont{Medvedeva}},
  \bibinfo{author}{\bibfnamefont{W.}~\bibnamefont{Wu}},
  \bibinfo{author}{\bibfnamefont{P.}~\bibnamefont{Ghaemi}},
  \bibinfo{author}{\bibfnamefont{T.}~\bibnamefont{Vojta}}, \bibnamefont{and}
  \bibinfo{author}{\bibfnamefont{Y.~S.} \bibnamefont{Hor}},
  \bibinfo{journal}{arXiv:1512.03519}  (\bibinfo{year}{2016}).

\bibitem[{\citenamefont{Lawson et~al.}(2016)\citenamefont{Lawson, Corbae, Li,
  Yu, Asaba, Tinsman, Qiu, Medvedeva, Hor, and Li}}]{PhysRevB.94.041114}
\bibinfo{author}{\bibfnamefont{B.~J.} \bibnamefont{Lawson}},
  \bibinfo{author}{\bibfnamefont{P.}~\bibnamefont{Corbae}},
  \bibinfo{author}{\bibfnamefont{G.}~\bibnamefont{Li}},
  \bibinfo{author}{\bibfnamefont{F.}~\bibnamefont{Yu}},
  \bibinfo{author}{\bibfnamefont{T.}~\bibnamefont{Asaba}},
  \bibinfo{author}{\bibfnamefont{C.}~\bibnamefont{Tinsman}},
  \bibinfo{author}{\bibfnamefont{Y.}~\bibnamefont{Qiu}},
  \bibinfo{author}{\bibfnamefont{J.~E.} \bibnamefont{Medvedeva}},
  \bibinfo{author}{\bibfnamefont{Y.~S.} \bibnamefont{Hor}}, \bibnamefont{and}
  \bibinfo{author}{\bibfnamefont{L.}~\bibnamefont{Li}}, \bibinfo{journal}{Phys.
  Rev. B} \textbf{\bibinfo{volume}{94}}, \bibinfo{pages}{041114}
  (\bibinfo{year}{2016}),
  \urlprefix\url{http://link.aps.org/doi/10.1103/PhysRevB.94.041114}.

\end{thebibliography}


\begin{thebibliography}{40}
\expandafter\ifx\csname natexlab\endcsname\relax\def\natexlab#1{#1}\fi
\expandafter\ifx\csname bibnamefont\endcsname\relax
  \def\bibnamefont#1{#1}\fi
\expandafter\ifx\csname bibfnamefont\endcsname\relax
  \def\bibfnamefont#1{#1}\fi
\expandafter\ifx\csname citenamefont\endcsname\relax
  \def\citenamefont#1{#1}\fi
\expandafter\ifx\csname url\endcsname\relax
  \def\url#1{\texttt{#1}}\fi
\expandafter\ifx\csname urlprefix\endcsname\relax\def\urlprefix{URL }\fi
\providecommand{\bibinfo}[2]{#2}
\providecommand{\eprint}[2][]{\url{#2}}

\bibitem[{\citenamefont{Sac\'{e}p\'{e}
  et~al.}(2011)\citenamefont{Sac\'{e}p\'{e}, Oostinga, Li, Ubaldini, Couto,
  Giannini, and Morpurgo}}]{NatCommun.2.575}
\bibinfo{author}{\bibfnamefont{B.}~\bibnamefont{Sac\'{e}p\'{e}}},
  \bibinfo{author}{\bibfnamefont{J.~B.} \bibnamefont{Oostinga}},
  \bibinfo{author}{\bibfnamefont{J.}~\bibnamefont{Li}},
  \bibinfo{author}{\bibfnamefont{A.}~\bibnamefont{Ubaldini}},
  \bibinfo{author}{\bibfnamefont{N.~J.} \bibnamefont{Couto}},
  \bibinfo{author}{\bibfnamefont{E.}~\bibnamefont{Giannini}}, \bibnamefont{and}
  \bibinfo{author}{\bibfnamefont{A.~F.} \bibnamefont{Morpurgo}},
  \bibinfo{journal}{Nat. Commun.} \textbf{\bibinfo{volume}{2}},
  \bibinfo{pages}{575} (\bibinfo{year}{2011}),
  \urlprefix\url{http://dx.doi.org/10.1038/ncomms1586}.

\bibitem[{\citenamefont{Williams et~al.}(2012)\citenamefont{Williams, Bestwick,
  Gallagher, Hong, Cui, Bleich, Analytis, Fisher, and
  Goldhaber-Gordon}}]{PhysRevLett.109.056803}
\bibinfo{author}{\bibfnamefont{J.~R.} \bibnamefont{Williams}},
  \bibinfo{author}{\bibfnamefont{A.~J.} \bibnamefont{Bestwick}},
  \bibinfo{author}{\bibfnamefont{P.}~\bibnamefont{Gallagher}},
  \bibinfo{author}{\bibfnamefont{S.~S.} \bibnamefont{Hong}},
  \bibinfo{author}{\bibfnamefont{Y.}~\bibnamefont{Cui}},
  \bibinfo{author}{\bibfnamefont{A.~S.} \bibnamefont{Bleich}},
  \bibinfo{author}{\bibfnamefont{J.~G.} \bibnamefont{Analytis}},
  \bibinfo{author}{\bibfnamefont{I.~R.} \bibnamefont{Fisher}},
  \bibnamefont{and}
  \bibinfo{author}{\bibfnamefont{D.}~\bibnamefont{Goldhaber-Gordon}},
  \bibinfo{journal}{Phys. Rev. Lett.} \textbf{\bibinfo{volume}{109}},
  \bibinfo{pages}{056803} (\bibinfo{year}{2012}),
  \urlprefix\url{http://link.aps.org/doi/10.1103/PhysRevLett.109.056803}.

\bibitem[{\citenamefont{Veldhorst et~al.}(2012)\citenamefont{Veldhorst,
  Snelder, Hoek, Gang, Guduru, Wang, Zeitler, van~der Wiel, Golubov, Hilgenkamp
  et~al.}}]{NatMat.11.417}
\bibinfo{author}{\bibfnamefont{M.}~\bibnamefont{Veldhorst}},
  \bibinfo{author}{\bibfnamefont{M.}~\bibnamefont{Snelder}},
  \bibinfo{author}{\bibfnamefont{M.}~\bibnamefont{Hoek}},
  \bibinfo{author}{\bibfnamefont{T.}~\bibnamefont{Gang}},
  \bibinfo{author}{\bibfnamefont{V.~K.} \bibnamefont{Guduru}},
  \bibinfo{author}{\bibfnamefont{X.~L.} \bibnamefont{Wang}},
  \bibinfo{author}{\bibfnamefont{U.}~\bibnamefont{Zeitler}},
  \bibinfo{author}{\bibfnamefont{W.~G.} \bibnamefont{van~der Wiel}},
  \bibinfo{author}{\bibfnamefont{A.~A.} \bibnamefont{Golubov}},
  \bibinfo{author}{\bibfnamefont{H.}~\bibnamefont{Hilgenkamp}},
  \bibnamefont{et~al.}, \bibinfo{journal}{Nat. Mater.}
  \textbf{\bibinfo{volume}{11}}, \bibinfo{pages}{417} (\bibinfo{year}{2012}),
  \urlprefix\url{dx.doi.org/10.1038/nmat3255}.

\bibitem[{\citenamefont{Cho et~al.}(2013)\citenamefont{Cho, Dellabetta, Yang,
  Schneeloch, Xu, Valla, Gu, Gilbert, and Mason}}]{NatCommun.4.1689}
\bibinfo{author}{\bibfnamefont{S.}~\bibnamefont{Cho}},
  \bibinfo{author}{\bibfnamefont{B.}~\bibnamefont{Dellabetta}},
  \bibinfo{author}{\bibfnamefont{A.}~\bibnamefont{Yang}},
  \bibinfo{author}{\bibfnamefont{J.}~\bibnamefont{Schneeloch}},
  \bibinfo{author}{\bibfnamefont{Z.}~\bibnamefont{Xu}},
  \bibinfo{author}{\bibfnamefont{T.}~\bibnamefont{Valla}},
  \bibinfo{author}{\bibfnamefont{G.}~\bibnamefont{Gu}},
  \bibinfo{author}{\bibfnamefont{M.~J.} \bibnamefont{Gilbert}},
  \bibnamefont{and} \bibinfo{author}{\bibfnamefont{N.}~\bibnamefont{Mason}},
  \bibinfo{journal}{Nat. Commun.} \textbf{\bibinfo{volume}{4}},
  \bibinfo{pages}{1689} (\bibinfo{year}{2013}),
  \urlprefix\url{http://dx.doi.org/10.1038/ncomms2701}.

\bibitem[{\citenamefont{Kurter et~al.}(2014)\citenamefont{Kurter, Finck,
  Ghaemi, Hor, and Van~Harlingen}}]{PhysRevB.90.014501}
\bibinfo{author}{\bibfnamefont{C.}~\bibnamefont{Kurter}},
  \bibinfo{author}{\bibfnamefont{A.~D.~K.} \bibnamefont{Finck}},
  \bibinfo{author}{\bibfnamefont{P.}~\bibnamefont{Ghaemi}},
  \bibinfo{author}{\bibfnamefont{Y.~S.} \bibnamefont{Hor}}, \bibnamefont{and}
  \bibinfo{author}{\bibfnamefont{D.~J.} \bibnamefont{Van~Harlingen}},
  \bibinfo{journal}{Phys. Rev. B} \textbf{\bibinfo{volume}{90}},
  \bibinfo{pages}{014501} (\bibinfo{year}{2014}),
  \urlprefix\url{http://link.aps.org/doi/10.1103/PhysRevB.90.014501}.

\bibitem[{\citenamefont{Stehno et~al.}(2016)\citenamefont{Stehno, Orlyanchik,
  Nugroho, Ghaemi, Brahlek, Koirala, Oh, and
  Van~Harlingen}}]{PhysRevB.93.035307}
\bibinfo{author}{\bibfnamefont{M.~P.} \bibnamefont{Stehno}},
  \bibinfo{author}{\bibfnamefont{V.}~\bibnamefont{Orlyanchik}},
  \bibinfo{author}{\bibfnamefont{C.~D.} \bibnamefont{Nugroho}},
  \bibinfo{author}{\bibfnamefont{P.}~\bibnamefont{Ghaemi}},
  \bibinfo{author}{\bibfnamefont{M.}~\bibnamefont{Brahlek}},
  \bibinfo{author}{\bibfnamefont{N.}~\bibnamefont{Koirala}},
  \bibinfo{author}{\bibfnamefont{S.}~\bibnamefont{Oh}}, \bibnamefont{and}
  \bibinfo{author}{\bibfnamefont{D.~J.} \bibnamefont{Van~Harlingen}},
  \bibinfo{journal}{Phys. Rev. B} \textbf{\bibinfo{volume}{93}},
  \bibinfo{pages}{035307} (\bibinfo{year}{2016}),
  \urlprefix\url{http://link.aps.org/doi/10.1103/PhysRevB.93.035307}.

\bibitem[{\citenamefont{Kurter et~al.}(2015)\citenamefont{Kurter, Finck, Hor,
  and Van~Harlingen}}]{KurterNature}
\bibinfo{author}{\bibfnamefont{C.}~\bibnamefont{Kurter}},
  \bibinfo{author}{\bibfnamefont{A.~D.~K.} \bibnamefont{Finck}},
  \bibinfo{author}{\bibfnamefont{Y.~S.} \bibnamefont{Hor}}, \bibnamefont{and}
  \bibinfo{author}{\bibfnamefont{D.~J.} \bibnamefont{Van~Harlingen}},
  \bibinfo{journal}{Nat. Commun.} \textbf{\bibinfo{volume}{6}},
  \bibinfo{pages}{7130} (\bibinfo{year}{2015}).

\bibitem[{\citenamefont{Hor et~al.}(2010)\citenamefont{Hor, Williams,
  Checkelsky, Roushan, Seo, Xu, Zandbergen, Yazdani, Ong, and
  Cava}}]{PhysRevLett.104.057001}
\bibinfo{author}{\bibfnamefont{Y.~S.} \bibnamefont{Hor}},
  \bibinfo{author}{\bibfnamefont{A.~J.} \bibnamefont{Williams}},
  \bibinfo{author}{\bibfnamefont{J.~G.} \bibnamefont{Checkelsky}},
  \bibinfo{author}{\bibfnamefont{P.}~\bibnamefont{Roushan}},
  \bibinfo{author}{\bibfnamefont{J.}~\bibnamefont{Seo}},
  \bibinfo{author}{\bibfnamefont{Q.}~\bibnamefont{Xu}},
  \bibinfo{author}{\bibfnamefont{H.~W.} \bibnamefont{Zandbergen}},
  \bibinfo{author}{\bibfnamefont{A.}~\bibnamefont{Yazdani}},
  \bibinfo{author}{\bibfnamefont{N.~P.} \bibnamefont{Ong}}, \bibnamefont{and}
  \bibinfo{author}{\bibfnamefont{R.~J.} \bibnamefont{Cava}},
  \bibinfo{journal}{Phys. Rev. Lett.} \textbf{\bibinfo{volume}{104}},
  \bibinfo{pages}{057001} (\bibinfo{year}{2010}),
  \urlprefix\url{http://link.aps.org/doi/10.1103/PhysRevLett.104.057001}.

\bibitem[{\citenamefont{Sasaki et~al.}(2011)\citenamefont{Sasaki, Kriener,
  Segawa, Yada, Tanaka, Sato, and Ando}}]{PhysRevLett.107.217001}
\bibinfo{author}{\bibfnamefont{S.}~\bibnamefont{Sasaki}},
  \bibinfo{author}{\bibfnamefont{M.}~\bibnamefont{Kriener}},
  \bibinfo{author}{\bibfnamefont{K.}~\bibnamefont{Segawa}},
  \bibinfo{author}{\bibfnamefont{K.}~\bibnamefont{Yada}},
  \bibinfo{author}{\bibfnamefont{Y.}~\bibnamefont{Tanaka}},
  \bibinfo{author}{\bibfnamefont{M.}~\bibnamefont{Sato}}, \bibnamefont{and}
  \bibinfo{author}{\bibfnamefont{Y.}~\bibnamefont{Ando}},
  \bibinfo{journal}{Phys. Rev. Lett.} \textbf{\bibinfo{volume}{107}},
  \bibinfo{pages}{217001} (\bibinfo{year}{2011}),
  \urlprefix\url{http://link.aps.org/doi/10.1103/PhysRevLett.107.217001}.

\bibitem[{\citenamefont{Qiu et~al.}(2016)\citenamefont{Qiu, Sanders, Dai,
  Medvedeva, Wu, Ghaemi, Vojta, and Hor}}]{YewSanArXIV}
\bibinfo{author}{\bibfnamefont{Y.}~\bibnamefont{Qiu}},
  \bibinfo{author}{\bibfnamefont{K.~N.} \bibnamefont{Sanders}},
  \bibinfo{author}{\bibfnamefont{J.}~\bibnamefont{Dai}},
  \bibinfo{author}{\bibfnamefont{J.~E.} \bibnamefont{Medvedeva}},
  \bibinfo{author}{\bibfnamefont{W.}~\bibnamefont{Wu}},
  \bibinfo{author}{\bibfnamefont{P.}~\bibnamefont{Ghaemi}},
  \bibinfo{author}{\bibfnamefont{T.}~\bibnamefont{Vojta}}, \bibnamefont{and}
  \bibinfo{author}{\bibfnamefont{Y.~S.} \bibnamefont{Hor}},
  \bibinfo{journal}{arXiv:1512.03519}  (\bibinfo{year}{2016}).

\bibitem[{\citenamefont{Liu et~al.}(2015)\citenamefont{Liu, Yao, Shao, Zuo, Pi,
  Tan, Zhang, and Zhang}}]{doi:10.1021/jacs.5b06815}
\bibinfo{author}{\bibfnamefont{Z.}~\bibnamefont{Liu}},
  \bibinfo{author}{\bibfnamefont{X.}~\bibnamefont{Yao}},
  \bibinfo{author}{\bibfnamefont{J.}~\bibnamefont{Shao}},
  \bibinfo{author}{\bibfnamefont{M.}~\bibnamefont{Zuo}},
  \bibinfo{author}{\bibfnamefont{L.}~\bibnamefont{Pi}},
  \bibinfo{author}{\bibfnamefont{S.}~\bibnamefont{Tan}},
  \bibinfo{author}{\bibfnamefont{C.}~\bibnamefont{Zhang}}, \bibnamefont{and}
  \bibinfo{author}{\bibfnamefont{Y.}~\bibnamefont{Zhang}},
  \bibinfo{journal}{Journal of the American Chemical Society}
  \textbf{\bibinfo{volume}{137}}, \bibinfo{pages}{10512}
  (\bibinfo{year}{2015}), \bibinfo{note}{pMID: 26262431},
  \eprint{http://dx.doi.org/10.1021/jacs.5b06815},
  \urlprefix\url{http://dx.doi.org/10.1021/jacs.5b06815}.

\bibitem[{\citenamefont{Peng et~al.}(2013)\citenamefont{Peng, De, Lv, Wei, and
  Chu}}]{PhysRevB.88.024515}
\bibinfo{author}{\bibfnamefont{H.}~\bibnamefont{Peng}},
  \bibinfo{author}{\bibfnamefont{D.}~\bibnamefont{De}},
  \bibinfo{author}{\bibfnamefont{B.}~\bibnamefont{Lv}},
  \bibinfo{author}{\bibfnamefont{F.}~\bibnamefont{Wei}}, \bibnamefont{and}
  \bibinfo{author}{\bibfnamefont{C.-W.} \bibnamefont{Chu}},
  \bibinfo{journal}{Phys. Rev. B} \textbf{\bibinfo{volume}{88}},
  \bibinfo{pages}{024515} (\bibinfo{year}{2013}),
  \urlprefix\url{http://link.aps.org/doi/10.1103/PhysRevB.88.024515}.

\bibitem[{\citenamefont{Hsieh and Fu}(2012)}]{PhysRevLett.108.107005}
\bibinfo{author}{\bibfnamefont{T.~H.} \bibnamefont{Hsieh}} \bibnamefont{and}
  \bibinfo{author}{\bibfnamefont{L.}~\bibnamefont{Fu}}, \bibinfo{journal}{Phys.
  Rev. Lett.} \textbf{\bibinfo{volume}{108}}, \bibinfo{pages}{107005}
  (\bibinfo{year}{2012}),
  \urlprefix\url{http://link.aps.org/doi/10.1103/PhysRevLett.108.107005}.

\bibitem[{\citenamefont{Majorana}(1937)}]{Majorana}
\bibinfo{author}{\bibfnamefont{E.}~\bibnamefont{Majorana}},
  \bibinfo{journal}{Nuovo Cimento} \textbf{\bibinfo{volume}{14}},
  \bibinfo{pages}{171} (\bibinfo{year}{1937}).

\bibitem[{\citenamefont{Wilczek}(2009)}]{NatPhys.5.614}
\bibinfo{author}{\bibfnamefont{F.}~\bibnamefont{Wilczek}},
  \bibinfo{journal}{Nat. Phys.} \textbf{\bibinfo{volume}{5}},
  \bibinfo{pages}{614} (\bibinfo{year}{2009}).

\bibitem[{\citenamefont{Nayak et~al.}(2008)\citenamefont{Nayak, Simon, Stern,
  Freedman, and Das~Sarma}}]{RevModPhys.80.1083}
\bibinfo{author}{\bibfnamefont{C.}~\bibnamefont{Nayak}},
  \bibinfo{author}{\bibfnamefont{S.~H.} \bibnamefont{Simon}},
  \bibinfo{author}{\bibfnamefont{A.}~\bibnamefont{Stern}},
  \bibinfo{author}{\bibfnamefont{M.}~\bibnamefont{Freedman}}, \bibnamefont{and}
  \bibinfo{author}{\bibfnamefont{S.}~\bibnamefont{Das~Sarma}},
  \bibinfo{journal}{Rev. Mod. Phys.} \textbf{\bibinfo{volume}{80}},
  \bibinfo{pages}{1083} (\bibinfo{year}{2008}).

\bibitem[{\citenamefont{Tinkham}(1996)}]{Tinkham}
\bibinfo{author}{\bibfnamefont{M.}~\bibnamefont{Tinkham}},
  \emph{\bibinfo{title}{Introduction to Superconductivity}}
  (\bibinfo{publisher}{Dover Publications, Inc.}, \bibinfo{year}{1996}).

\bibitem[{\citenamefont{Levy et~al.}(2013)\citenamefont{Levy, Zhang, Ha,
  Sharifi, Talin, Kuk, and Stroscio}}]{PhysRevLett.110.117001}
\bibinfo{author}{\bibfnamefont{N.}~\bibnamefont{Levy}},
  \bibinfo{author}{\bibfnamefont{T.}~\bibnamefont{Zhang}},
  \bibinfo{author}{\bibfnamefont{J.}~\bibnamefont{Ha}},
  \bibinfo{author}{\bibfnamefont{F.}~\bibnamefont{Sharifi}},
  \bibinfo{author}{\bibfnamefont{A.~A.} \bibnamefont{Talin}},
  \bibinfo{author}{\bibfnamefont{Y.}~\bibnamefont{Kuk}}, \bibnamefont{and}
  \bibinfo{author}{\bibfnamefont{J.~A.} \bibnamefont{Stroscio}},
  \bibinfo{journal}{Phys. Rev. Lett.} \textbf{\bibinfo{volume}{110}},
  \bibinfo{pages}{117001} (\bibinfo{year}{2013}),
  \urlprefix\url{http://link.aps.org/doi/10.1103/PhysRevLett.110.117001}.

\bibitem[{\citenamefont{Matano et~al.}(2016)\citenamefont{Matano, Kriener,
  Segawa, Ando, and Zheng}}]{NatPhys12_852}
\bibinfo{author}{\bibfnamefont{K.}~\bibnamefont{Matano}},
  \bibinfo{author}{\bibfnamefont{M.}~\bibnamefont{Kriener}},
  \bibinfo{author}{\bibfnamefont{K.}~\bibnamefont{Segawa}},
  \bibinfo{author}{\bibfnamefont{Y.}~\bibnamefont{Ando}}, \bibnamefont{and}
  \bibinfo{author}{\bibfnamefont{G.}~\bibnamefont{Zheng}},
  \bibinfo{journal}{Nat. Phys.} \textbf{\bibinfo{volume}{12}},
  \bibinfo{pages}{852} (\bibinfo{year}{2016}).

\bibitem[{\citenamefont{Chirolli et~al.}(2016)\citenamefont{Chirolli, de~Juan,
  and Guinea}}]{ChirolliArXIV}
\bibinfo{author}{\bibfnamefont{L.}~\bibnamefont{Chirolli}},
  \bibinfo{author}{\bibfnamefont{F.}~\bibnamefont{de~Juan}}, \bibnamefont{and}
  \bibinfo{author}{\bibfnamefont{F.}~\bibnamefont{Guinea}},
  \bibinfo{journal}{arXiv:1611.02173v1}  (\bibinfo{year}{2016}).

\bibitem[{\citenamefont{Yuan et~al.}(2016)\citenamefont{Yuan, He, and
  Law}}]{LawArXIV}
\bibinfo{author}{\bibfnamefont{N.~F.~Q.} \bibnamefont{Yuan}},
  \bibinfo{author}{\bibfnamefont{W.-Y.} \bibnamefont{He}}, \bibnamefont{and}
  \bibinfo{author}{\bibfnamefont{K.~T.} \bibnamefont{Law}},
  \bibinfo{journal}{arXiv:1608.05825v2}  (\bibinfo{year}{2016}).

\bibitem[{\citenamefont{Asaba et~al.}(2017)\citenamefont{Asaba, Lawson,
  Tinsman, Chen, Corbae, Li, Qiu, Hor, Fu, and Li}}]{PhysRevX.7.011009}
\bibinfo{author}{\bibfnamefont{T.}~\bibnamefont{Asaba}},
  \bibinfo{author}{\bibfnamefont{B.~J.} \bibnamefont{Lawson}},
  \bibinfo{author}{\bibfnamefont{C.}~\bibnamefont{Tinsman}},
  \bibinfo{author}{\bibfnamefont{L.}~\bibnamefont{Chen}},
  \bibinfo{author}{\bibfnamefont{P.}~\bibnamefont{Corbae}},
  \bibinfo{author}{\bibfnamefont{G.}~\bibnamefont{Li}},
  \bibinfo{author}{\bibfnamefont{Y.}~\bibnamefont{Qiu}},
  \bibinfo{author}{\bibfnamefont{Y.~S.} \bibnamefont{Hor}},
  \bibinfo{author}{\bibfnamefont{L.}~\bibnamefont{Fu}}, \bibnamefont{and}
  \bibinfo{author}{\bibfnamefont{L.}~\bibnamefont{Li}}, \bibinfo{journal}{Phys.
  Rev. X} \textbf{\bibinfo{volume}{7}}, \bibinfo{pages}{011009}
  (\bibinfo{year}{2017}),
  \urlprefix\url{http://link.aps.org/doi/10.1103/PhysRevX.7.011009}.

\bibitem[{\citenamefont{Smylie et~al.}(2016)\citenamefont{Smylie, Claus, Welp,
  Kwok, Qiu, Hor, and Snezhko}}]{PhysRevB.94.180510}
\bibinfo{author}{\bibfnamefont{M.~P.} \bibnamefont{Smylie}},
  \bibinfo{author}{\bibfnamefont{H.}~\bibnamefont{Claus}},
  \bibinfo{author}{\bibfnamefont{U.}~\bibnamefont{Welp}},
  \bibinfo{author}{\bibfnamefont{W.-K.} \bibnamefont{Kwok}},
  \bibinfo{author}{\bibfnamefont{Y.}~\bibnamefont{Qiu}},
  \bibinfo{author}{\bibfnamefont{Y.~S.} \bibnamefont{Hor}}, \bibnamefont{and}
  \bibinfo{author}{\bibfnamefont{A.}~\bibnamefont{Snezhko}},
  \bibinfo{journal}{Phys. Rev. B} \textbf{\bibinfo{volume}{94}},
  \bibinfo{pages}{180510} (\bibinfo{year}{2016}),
  \urlprefix\url{http://link.aps.org/doi/10.1103/PhysRevB.94.180510}.

\bibitem[{\citenamefont{Blonder et~al.}(1982)\citenamefont{Blonder, Tinkham,
  and Klapwijk}}]{PhysRevB.25.4515}
\bibinfo{author}{\bibfnamefont{G.~E.} \bibnamefont{Blonder}},
  \bibinfo{author}{\bibfnamefont{M.}~\bibnamefont{Tinkham}}, \bibnamefont{and}
  \bibinfo{author}{\bibfnamefont{T.~M.} \bibnamefont{Klapwijk}},
  \bibinfo{journal}{Phys. Rev. B} \textbf{\bibinfo{volume}{25}},
  \bibinfo{pages}{4515} (\bibinfo{year}{1982}),
  \urlprefix\url{http://link.aps.org/doi/10.1103/PhysRevB.25.4515}.

\bibitem[{\citenamefont{Yamakage et~al.}(2012)\citenamefont{Yamakage, Yada,
  Sato, and Tanaka}}]{PhysRevB.85.180509}
\bibinfo{author}{\bibfnamefont{A.}~\bibnamefont{Yamakage}},
  \bibinfo{author}{\bibfnamefont{K.}~\bibnamefont{Yada}},
  \bibinfo{author}{\bibfnamefont{M.}~\bibnamefont{Sato}}, \bibnamefont{and}
  \bibinfo{author}{\bibfnamefont{Y.}~\bibnamefont{Tanaka}},
  \bibinfo{journal}{Phys. Rev. B} \textbf{\bibinfo{volume}{85}},
  \bibinfo{pages}{180509} (\bibinfo{year}{2012}),
  \urlprefix\url{http://link.aps.org/doi/10.1103/PhysRevB.85.180509}.

\bibitem[{\citenamefont{Sheet et~al.}(2004)\citenamefont{Sheet, Mukhopadhyay,
  and Raychaudhuri}}]{PhysRevB.69.134507}
\bibinfo{author}{\bibfnamefont{G.}~\bibnamefont{Sheet}},
  \bibinfo{author}{\bibfnamefont{S.}~\bibnamefont{Mukhopadhyay}},
  \bibnamefont{and}
  \bibinfo{author}{\bibfnamefont{P.}~\bibnamefont{Raychaudhuri}},
  \bibinfo{journal}{Phys. Rev. B} \textbf{\bibinfo{volume}{69}},
  \bibinfo{pages}{134507} (\bibinfo{year}{2004}),
  \urlprefix\url{https://link.aps.org/doi/10.1103/PhysRevB.69.134507}.

\bibitem[{\citenamefont{Yamashiro et~al.}(1997)\citenamefont{Yamashiro, Tanaka,
  and Kashiwaya}}]{PhysRevB.56.7847}
\bibinfo{author}{\bibfnamefont{M.}~\bibnamefont{Yamashiro}},
  \bibinfo{author}{\bibfnamefont{Y.}~\bibnamefont{Tanaka}}, \bibnamefont{and}
  \bibinfo{author}{\bibfnamefont{S.}~\bibnamefont{Kashiwaya}},
  \bibinfo{journal}{Phys. Rev. B} \textbf{\bibinfo{volume}{56}},
  \bibinfo{pages}{7847} (\bibinfo{year}{1997}),
  \urlprefix\url{http://link.aps.org/doi/10.1103/PhysRevB.56.7847}.

\bibitem[{\citenamefont{Yamashiro et~al.}(1998)\citenamefont{Yamashiro, Tanaka,
  Tanuma, and Kashiwaya}}]{KashiwayaJPSJ}
\bibinfo{author}{\bibfnamefont{M.}~\bibnamefont{Yamashiro}},
  \bibinfo{author}{\bibfnamefont{Y.}~\bibnamefont{Tanaka}},
  \bibinfo{author}{\bibfnamefont{Y.}~\bibnamefont{Tanuma}}, \bibnamefont{and}
  \bibinfo{author}{\bibfnamefont{S.}~\bibnamefont{Kashiwaya}},
  \bibinfo{journal}{Journal of the Physical Society of Japan}
  \textbf{\bibinfo{volume}{67}}, \bibinfo{pages}{3224} (\bibinfo{year}{1998}),
  \eprint{http://dx.doi.org/10.1143/JPSJ.67.3224},
  \urlprefix\url{http://dx.doi.org/10.1143/JPSJ.67.3224}.

\bibitem[{\citenamefont{Burset et~al.}(2015)\citenamefont{Burset, Lu, Tkachov,
  Tanaka, Hankiewicz, and Trauzettel}}]{PhysRevB.92.205424}
\bibinfo{author}{\bibfnamefont{P.}~\bibnamefont{Burset}},
  \bibinfo{author}{\bibfnamefont{B.}~\bibnamefont{Lu}},
  \bibinfo{author}{\bibfnamefont{G.}~\bibnamefont{Tkachov}},
  \bibinfo{author}{\bibfnamefont{Y.}~\bibnamefont{Tanaka}},
  \bibinfo{author}{\bibfnamefont{E.~M.} \bibnamefont{Hankiewicz}},
  \bibnamefont{and}
  \bibinfo{author}{\bibfnamefont{B.}~\bibnamefont{Trauzettel}},
  \bibinfo{journal}{Phys. Rev. B} \textbf{\bibinfo{volume}{92}},
  \bibinfo{pages}{205424} (\bibinfo{year}{2015}),
  \urlprefix\url{http://link.aps.org/doi/10.1103/PhysRevB.92.205424}.

\bibitem[{\citenamefont{Yuan et~al.}(2017)\citenamefont{Yuan, He, and
  Law}}]{PhysRevB.95.201109}
\bibinfo{author}{\bibfnamefont{N.~F.~Q.} \bibnamefont{Yuan}},
  \bibinfo{author}{\bibfnamefont{W.-Y.} \bibnamefont{He}}, \bibnamefont{and}
  \bibinfo{author}{\bibfnamefont{K.~T.} \bibnamefont{Law}},
  \bibinfo{journal}{Phys. Rev. B} \textbf{\bibinfo{volume}{95}},
  \bibinfo{pages}{201109} (\bibinfo{year}{2017}),
  \urlprefix\url{https://link.aps.org/doi/10.1103/PhysRevB.95.201109}.

\bibitem[{\citenamefont{Chirolli et~al.}(2017)\citenamefont{Chirolli, de~Juan,
  and Guinea}}]{PhysRevB.95.201110}
\bibinfo{author}{\bibfnamefont{L.}~\bibnamefont{Chirolli}},
  \bibinfo{author}{\bibfnamefont{F.}~\bibnamefont{de~Juan}}, \bibnamefont{and}
  \bibinfo{author}{\bibfnamefont{F.}~\bibnamefont{Guinea}},
  \bibinfo{journal}{Phys. Rev. B} \textbf{\bibinfo{volume}{95}},
  \bibinfo{pages}{201110} (\bibinfo{year}{2017}),
  \urlprefix\url{https://link.aps.org/doi/10.1103/PhysRevB.95.201110}.

\bibitem[{\citenamefont{Nandi et~al.}(2013)\citenamefont{Nandi, Khaire, Finck,
  Eisenstein, Pfeiffer, and West}}]{PhysRevB.88.165308}
\bibinfo{author}{\bibfnamefont{D.}~\bibnamefont{Nandi}},
  \bibinfo{author}{\bibfnamefont{T.}~\bibnamefont{Khaire}},
  \bibinfo{author}{\bibfnamefont{A.~D.~K.} \bibnamefont{Finck}},
  \bibinfo{author}{\bibfnamefont{J.~P.} \bibnamefont{Eisenstein}},
  \bibinfo{author}{\bibfnamefont{L.~N.} \bibnamefont{Pfeiffer}},
  \bibnamefont{and} \bibinfo{author}{\bibfnamefont{K.~W.} \bibnamefont{West}},
  \bibinfo{journal}{Phys. Rev. B} \textbf{\bibinfo{volume}{88}},
  \bibinfo{pages}{165308} (\bibinfo{year}{2013}),
  \urlprefix\url{http://link.aps.org/doi/10.1103/PhysRevB.88.165308}.

\bibitem[{\citenamefont{Kashiwaya et~al.}(2011)\citenamefont{Kashiwaya,
  Kashiwaya, Kambara, Furuta, Yaguchi, Tanaka, and
  Maeno}}]{PhysRevLett.107.077003}
\bibinfo{author}{\bibfnamefont{S.}~\bibnamefont{Kashiwaya}},
  \bibinfo{author}{\bibfnamefont{H.}~\bibnamefont{Kashiwaya}},
  \bibinfo{author}{\bibfnamefont{H.}~\bibnamefont{Kambara}},
  \bibinfo{author}{\bibfnamefont{T.}~\bibnamefont{Furuta}},
  \bibinfo{author}{\bibfnamefont{H.}~\bibnamefont{Yaguchi}},
  \bibinfo{author}{\bibfnamefont{Y.}~\bibnamefont{Tanaka}}, \bibnamefont{and}
  \bibinfo{author}{\bibfnamefont{Y.}~\bibnamefont{Maeno}},
  \bibinfo{journal}{Phys. Rev. Lett.} \textbf{\bibinfo{volume}{107}},
  \bibinfo{pages}{077003} (\bibinfo{year}{2011}),
  \urlprefix\url{http://link.aps.org/doi/10.1103/PhysRevLett.107.077003}.

\bibitem[{\citenamefont{Lawson et~al.}(2016)\citenamefont{Lawson, Corbae, Li,
  Yu, Asaba, Tinsman, Qiu, Medvedeva, Hor, and Li}}]{PhysRevB.94.041114}
\bibinfo{author}{\bibfnamefont{B.~J.} \bibnamefont{Lawson}},
  \bibinfo{author}{\bibfnamefont{P.}~\bibnamefont{Corbae}},
  \bibinfo{author}{\bibfnamefont{G.}~\bibnamefont{Li}},
  \bibinfo{author}{\bibfnamefont{F.}~\bibnamefont{Yu}},
  \bibinfo{author}{\bibfnamefont{T.}~\bibnamefont{Asaba}},
  \bibinfo{author}{\bibfnamefont{C.}~\bibnamefont{Tinsman}},
  \bibinfo{author}{\bibfnamefont{Y.}~\bibnamefont{Qiu}},
  \bibinfo{author}{\bibfnamefont{J.~E.} \bibnamefont{Medvedeva}},
  \bibinfo{author}{\bibfnamefont{Y.~S.} \bibnamefont{Hor}}, \bibnamefont{and}
  \bibinfo{author}{\bibfnamefont{L.}~\bibnamefont{Li}}, \bibinfo{journal}{Phys.
  Rev. B} \textbf{\bibinfo{volume}{94}}, \bibinfo{pages}{041114}
  (\bibinfo{year}{2016}),
  \urlprefix\url{http://link.aps.org/doi/10.1103/PhysRevB.94.041114}.

\bibitem[{\citenamefont{Cherkez et~al.}(2014)\citenamefont{Cherkez, Cuevas,
  Brun, Cren, M\'enard, Debontridder, Stolyarov, and
  Roditchev}}]{PhysRevX.4.011033}
\bibinfo{author}{\bibfnamefont{V.}~\bibnamefont{Cherkez}},
  \bibinfo{author}{\bibfnamefont{J.~C.} \bibnamefont{Cuevas}},
  \bibinfo{author}{\bibfnamefont{C.}~\bibnamefont{Brun}},
  \bibinfo{author}{\bibfnamefont{T.}~\bibnamefont{Cren}},
  \bibinfo{author}{\bibfnamefont{G.}~\bibnamefont{M\'enard}},
  \bibinfo{author}{\bibfnamefont{F.}~\bibnamefont{Debontridder}},
  \bibinfo{author}{\bibfnamefont{V.~S.} \bibnamefont{Stolyarov}},
  \bibnamefont{and}
  \bibinfo{author}{\bibfnamefont{D.}~\bibnamefont{Roditchev}},
  \bibinfo{journal}{Phys. Rev. X} \textbf{\bibinfo{volume}{4}},
  \bibinfo{pages}{011033} (\bibinfo{year}{2014}),
  \urlprefix\url{https://link.aps.org/doi/10.1103/PhysRevX.4.011033}.

\bibitem[{\citenamefont{Goldhaber-Gordon
  et~al.}(1998)\citenamefont{Goldhaber-Gordon, Shtrikman, Mahalu,
  Abusch-Magder, Meirav, and Kastner}}]{nature.391.156}
\bibinfo{author}{\bibfnamefont{D.}~\bibnamefont{Goldhaber-Gordon}},
  \bibinfo{author}{\bibfnamefont{H.}~\bibnamefont{Shtrikman}},
  \bibinfo{author}{\bibfnamefont{D.}~\bibnamefont{Mahalu}},
  \bibinfo{author}{\bibfnamefont{D.}~\bibnamefont{Abusch-Magder}},
  \bibinfo{author}{\bibfnamefont{U.}~\bibnamefont{Meirav}}, \bibnamefont{and}
  \bibinfo{author}{\bibfnamefont{M.~A.} \bibnamefont{Kastner}},
  \bibinfo{journal}{Nature} \textbf{\bibinfo{volume}{391}},
  \bibinfo{pages}{156} (\bibinfo{year}{1998}).

\bibitem[{\citenamefont{Kim et~al.}(1965)\citenamefont{Kim, Hempstead, and
  Strnad}}]{PhysRev.139.A1163}
\bibinfo{author}{\bibfnamefont{Y.~B.} \bibnamefont{Kim}},
  \bibinfo{author}{\bibfnamefont{C.~F.} \bibnamefont{Hempstead}},
  \bibnamefont{and} \bibinfo{author}{\bibfnamefont{A.~R.}
  \bibnamefont{Strnad}}, \bibinfo{journal}{Phys. Rev.}
  \textbf{\bibinfo{volume}{139}}, \bibinfo{pages}{A1163}
  (\bibinfo{year}{1965}),
  \urlprefix\url{http://link.aps.org/doi/10.1103/PhysRev.139.A1163}.

\bibitem[{\citenamefont{Geim et~al.}(1997)\citenamefont{Geim, Grigorieva,
  Dubonos, Lok, Maan, Filippov, and Peeters}}]{Nature.390.259}
\bibinfo{author}{\bibfnamefont{A.~K.} \bibnamefont{Geim}},
  \bibinfo{author}{\bibfnamefont{I.~V.} \bibnamefont{Grigorieva}},
  \bibinfo{author}{\bibfnamefont{S.~V.} \bibnamefont{Dubonos}},
  \bibinfo{author}{\bibfnamefont{J.~G.~S.} \bibnamefont{Lok}},
  \bibinfo{author}{\bibfnamefont{J.~C.} \bibnamefont{Maan}},
  \bibinfo{author}{\bibfnamefont{A.~E.} \bibnamefont{Filippov}},
  \bibnamefont{and} \bibinfo{author}{\bibfnamefont{F.~M.}
  \bibnamefont{Peeters}}, \bibinfo{journal}{Nature}
  \textbf{\bibinfo{volume}{390}}, \bibinfo{pages}{259} (\bibinfo{year}{1997}),
  \urlprefix\url{http://dx.doi.org/10.1038/36797}.

\bibitem[{\citenamefont{Fu and Kane}(2008)}]{PhysRevLett.100.096407}
\bibinfo{author}{\bibfnamefont{L.}~\bibnamefont{Fu}} \bibnamefont{and}
  \bibinfo{author}{\bibfnamefont{C.~L.} \bibnamefont{Kane}},
  \bibinfo{journal}{Phys. Rev. Lett.} \textbf{\bibinfo{volume}{100}},
  \bibinfo{pages}{096407} (\bibinfo{year}{2008}),
  \urlprefix\url{https://link.aps.org/doi/10.1103/PhysRevLett.100.096407}.

\bibitem[{\citenamefont{Hosur et~al.}(2011)\citenamefont{Hosur, Ghaemi, Mong,
  and Vishwanath}}]{PhysRevLett.107.097001}
\bibinfo{author}{\bibfnamefont{P.}~\bibnamefont{Hosur}},
  \bibinfo{author}{\bibfnamefont{P.}~\bibnamefont{Ghaemi}},
  \bibinfo{author}{\bibfnamefont{R.~S.~K.} \bibnamefont{Mong}},
  \bibnamefont{and}
  \bibinfo{author}{\bibfnamefont{A.}~\bibnamefont{Vishwanath}},
  \bibinfo{journal}{Phys. Rev. Lett.} \textbf{\bibinfo{volume}{107}},
  \bibinfo{pages}{097001} (\bibinfo{year}{2011}),
  \urlprefix\url{https://link.aps.org/doi/10.1103/PhysRevLett.107.097001}.

\end{thebibliography}

\end{document}


\title{{Supplemental Material: Andreev Reflection Spectroscopy of Topological Superconductor Candidate Nb$_x$Bi$_2$Se$_3$}}

\author{C.~Kurter}
\affiliation{Department of Physics and Materials Research Center, Missouri University of Science and Technology, Rolla, MO 65409}
\affiliation{Department of Physics and Materials Research Laboratory, University of Illinois at Urbana-Champaign,
Urbana, IL 61801}

\author{A.~D.~K.~Finck}
\affiliation{Department of Physics and Materials Research Laboratory, University of Illinois at Urbana-Champaign,
Urbana, IL 61801}

\author{E.~D.~Huemiller}
\affiliation{Department of Physics and Materials Research Laboratory, University of Illinois at Urbana-Champaign,
Urbana, IL 61801}

\author{J.~Medvedeva}
\affiliation{Department of Physics and Materials Research Center, Missouri University of Science and Technology, Rolla, MO 65409}

\author{A.~Weis}
\affiliation{Department of Physics and Materials Research Laboratory, University of Illinois at Urbana-Champaign,
Urbana, IL 61801}

\author{J.~M.~Atkinson}
\affiliation{Department of Physics and Materials Research Laboratory, University of Illinois at Urbana-Champaign,
Urbana, IL 61801}

\author{Y.~Qiu}
\affiliation{Department of Physics and Materials Research Center, Missouri University of Science and Technology, Rolla, MO 65409}
\author{L.~Shen}
\affiliation{Department of Physics and Materials Research Center, Missouri University of Science and Technology, Rolla, MO 65409}
\author{S.~H.~Lee}
\affiliation{Department of Physics and Materials Research Center, Missouri University of Science and Technology, Rolla, MO 65409}
\author{T.~Vojta}
\affiliation{Department of Physics and Materials Research Center, Missouri University of Science and Technology, Rolla, MO 65409}

\author{P. Ghaemi}
\affiliation{Department of Physics, City College of New of CUNY, New York, NY 10031}
\affiliation{Department of Physics, Graduate Center of CUNY, New York, NY 10016}

\author{Y.~S.~Hor}
\affiliation{Department of Physics and Materials Research Center, Missouri University of Science and Technology, Rolla, MO 65409}

\author{D.~J.~Van Harlingen}
\affiliation{Department of Physics and Materials Research Laboratory, University of Illinois at Urbana-Champaign,
Urbana, IL 61801}

\maketitle

\section{I. Zero bias conductance}

\begin{figure}
\centering
\includegraphics[bb=5 5 830 370,width= 6 in]{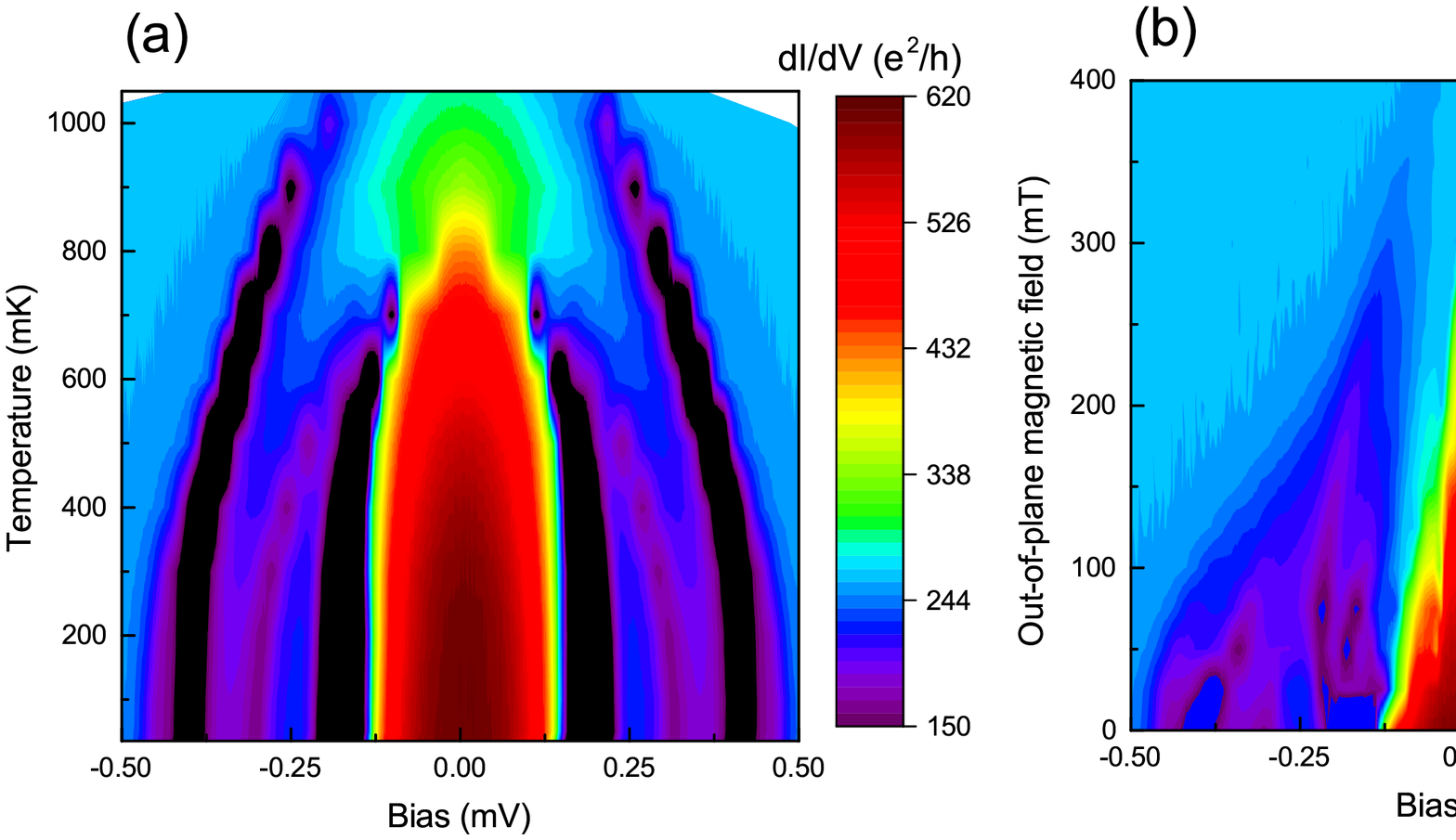}
\caption{(Color online) Temperature (a) and magnetic field (b) dependence of conductance spectra from a device with transparent contacts. Zero bias conductance peaks gradually get weaker with both temperature and magnetic field. } \label{fig:supp_ZBC}
\end{figure}

A superconducting device can be treated as a circuit consisting of a resistive element and a superconducting (zero voltage) element. For such a circuit, a finite critical current can generate the appearance of a zero bias conductance peak. When the bias current exceeds the critical current of the superconductor, the device becomes resistive and increases the total circuit resistance. In this picture, the zero bias resistance is equal to that of the resistive element in series with the superconductor, such as contact resistance.  We do not expect contact resistance to be strongly dependent on temperature or magnetic field, although the critical current is expected to vary with respect to either such variable.  Thus, one would expect the width of the apparent zero bias conductance peak to narrow with increased temperature or magnetic field, but its height should be essentially unchanged.

In Fig.~S1, we show color plots of dI/dV vs V as a function of temperature (a) and magnetic field (b) from an Andreev reflection device made with 16 nm thick flake with low contact resistance.  We find that zero bias conductance drops steadily with finite temperature or magnetic field, even as the conductance dips at finite bias remain visible.  This is in contrast with the simple picture of the zero bias conductance peak originating from finite critical current and supports the hypothesis that the peak is a signature of low energy bound states~\cite{PhysRevLett.107.217001}.

\section{II. Electronic Band Calculations}

\begin{figure}
\centering
\includegraphics[bb=0 0 649 360,width= 5 in]{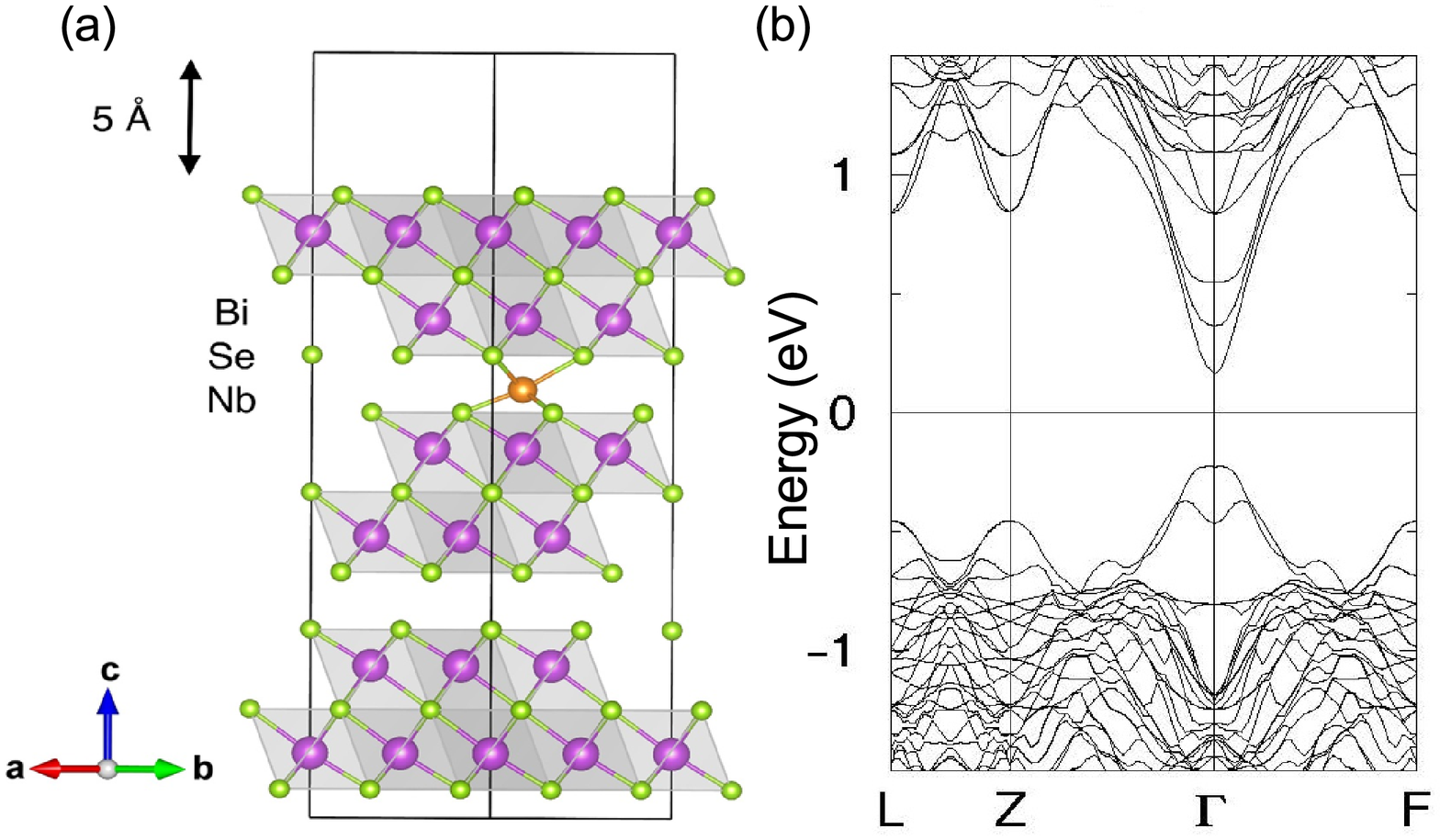}
\caption{(Color online) (a) The slab used in the calculations showing quintuple layers of Bi$_2$Se$_3$ and single intercalated Nb atom. (b) Calculated electronic structure of the Bi$_2$Se$_3$ slab.} \label{fig:crystal}
\end{figure}

The observation of multiple gaps in some of the measured differential conductance spectra (see Fig.~3) led us to consider that these spectral features might be due to the existence of multi-band superconductivity. We calculated band structures of Bi$_2$Se$_3$ and Nb-intercalated Bi$_2$Se$_3$ slabs by using the generalized gradient approximation in the Perdew-Burke-Ernzerhof form~\cite{PhysRevLett.77.3865, PhysRevLett.78.1396}. Figure~\ref{fig:crystal}(a) shows the electronic structure used in the calculations. To study 2D structures under periodic boundary conditions, a vacuum layer with a thickness of 5 \AA~was inserted to avoid the interaction between the periodic slabs of $2 \times 2$ rhombohedral Bi$_2$Se$_3$ with three quintuple layers. A single Nb atom (shown in orange) was added either at the surface, inside the top quintuple layer, or between the top two quintuple layers. The energetically preferred location of a single Nb atom within the Bi$_2$Se$_3$ slab was determined using the total energy calculations. Similar to the previous investigations for Nb-intercalated bulk Bi$_2$Se$_3$~\cite{YewSanArXIV}, it is found that Nb prefers to be located between the quintuple layers of the Bi$_2$Se$_3$ slab. For comparison, we first calculated the electronic band structure of Bi$_2$Se$_3$ slab without Nb intercalation as shown in Fig.~\ref{fig:crystal}(b).

\begin{figure}
\centering
\includegraphics[bb=0 0 649 360,width= 5 in]{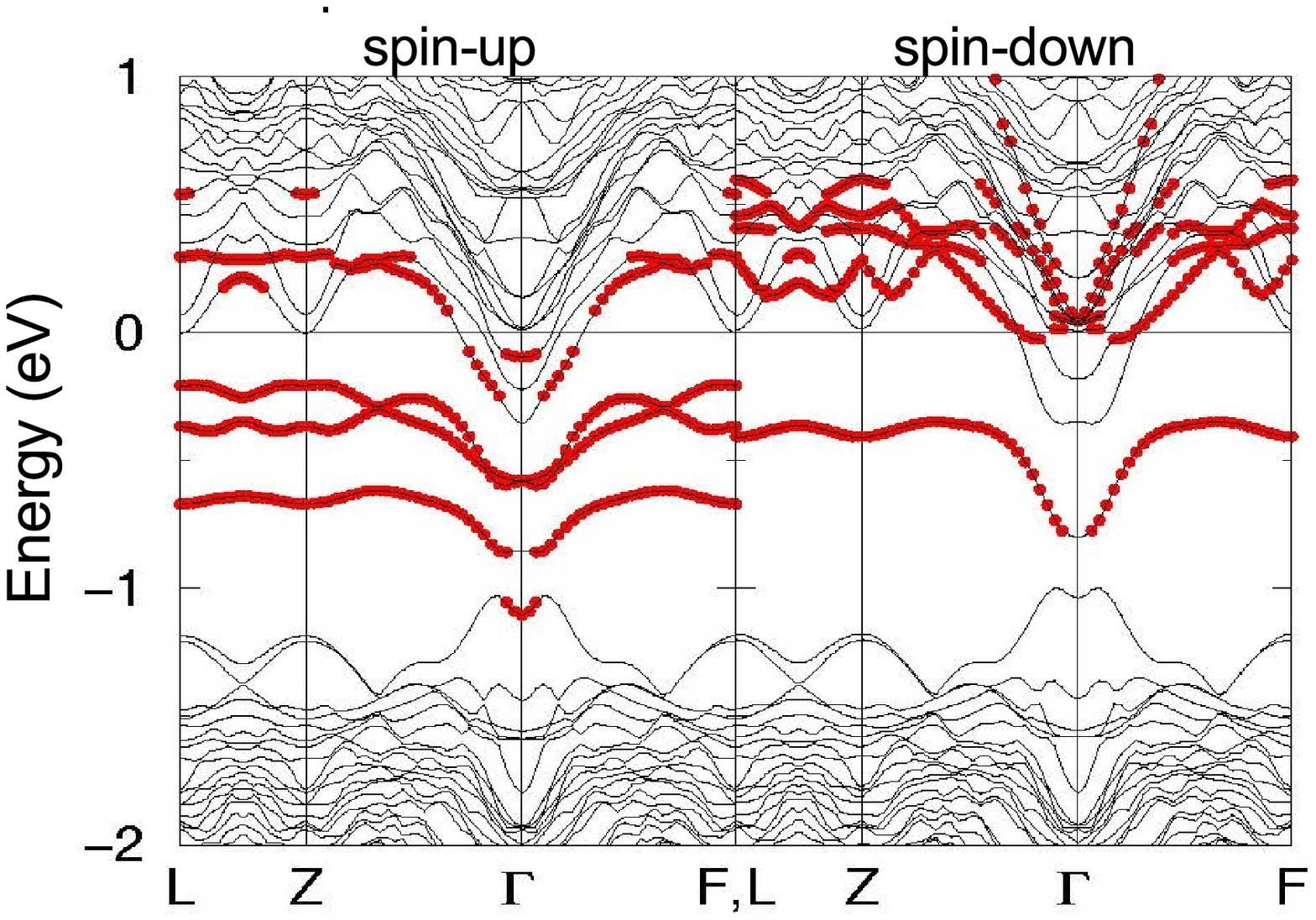}
\caption{(Color online) Comparison of spin up and spin down case of the calculated electronic structure of the Bi$_2$Se$_3$ slab with Nb intercalation.} \label{fig:supp_DFT}
\end{figure}

Intoduction of Nb into the slab dramatically changes the band structure as shown in Fig.~\ref{fig:supp_DFT}. The occupied Nb states appearing within the band gap of Bi$_2$Se$_3$ below the Fermi energy are highlighted in red; one can see three states for spin-up case and one state for spin-down case. More importantly, Nb states hybridize with the host Bi$_2$Se$_3$ states near the Fermi level resulting in a spin-dependent band structure. The spin-resolved differences near Fermi level suggest previously observed multiple Fermi surfaces~\cite{PhysRevB.94.041114} for Nb-intercalated Bi$_2$Se$_3$, which supports our experimental observations of Andreev conductance spectra with multiple gap features.

\bibliography{NbxBi2Se3supp}